# Tools and Procedures for the CTA Array Calibration

**Maria Concetta Maccarone**[1]
*Istituto Nazionale di Astrofisica, INAF – IASF Palermo, Via Ugo La Malfa 153, I-90146 Palermo, Italy*
*E-mail:* `Cettina.Maccarone@iasf-palermo.inaf.it`

**Dan Parsons**
*Max-Planck-Institut für Kernphysik, P.O. Box 103980, D-69029 Heidelberg, Germany*
*E-mail:* `Daniel.Parsons@mpi-hd.mpg.de`

**Markus Gaug**
*Universitat Autònoma de Barcelona, Departament de Física, E-08193 Bellaterra, Spain*
*E-mail:* `Markus.Gaug@uab.cat`

**Raquel de los Reyes**[*]
*Max-Planck-Institut für Kernphysik, P.O. Box 103980, D-69029 Heidelberg, Germany*
*E-mail:* `Reyes@mpi-hd.mpg.de`

**for the CTA Consortium**
`http://cta-observatory.org`

The Cherenkov Telescope Array (CTA) is an international initiative to build the next generation ground-based very-high-energy gamma-ray observatory. Full sky coverage will be assured by two arrays, one located on each of the northern and southern hemispheres. Three different sizes of telescopes will cover a wide energy range from tens of GeV up to hundreds of TeV. These telescopes, of which prototypes are currently under construction or completion, will have different mirror sizes and fields-of-view designed to access different energy regimes. Additionally, there will be groups of telescopes with different optics system, camera and electronics design. Given this diversity of instruments, an overall coherent calibration of the full array is a challenging task. Moreover, the CTA's requirements on calibration accuracy are much more stringent than those achieved with current Imaging Atmospheric Cherenkov Telescopes, like for instance: the systematic errors in the energy scale must not exceed 10%. In this contribution we present both the methods that, applied directly to the acquired observational CTA data, will ensure that the calibration is correctly performed to the stringent required precision, and the calibration equipment that, external to the telescopes, is currently under development and testing. Moreover, some notes about the operative procedure to be followed with both methods and instruments, will be described. The methods applied to the observational CTA data include the analysis of muon ring images, of carefully selected cosmic-ray air shower images, of the reconstructed electron spectrum and that of known gamma-ray sources and the possible use of stereo techniques hardware-independent. These methods will be complemented with the use of calibrated light sources located on ground or on board unmanned aerial vehicles.

---

[1]Speaker
[*] *Now with the German Aerospace Center (DLR), Earth Observation Center, 82234 Wessling, Germany*

## 1. Introduction

The Cherenkov Telescope Array (CTA) [1, 2] is the next generation ground-based observatory for gamma-ray astronomy at very high energies (VHE). With more than one hundred Cherenkov telescopes located at two sites, one for each hemisphere, assuring full sky coverage, CTA will detect VHE radiation with unprecedented accuracy and a sensitivity that is approximately 10 times better than current Imaging Atmospheric Cherenkov Technique (IACT) experiments. Three different classes of telescopes (LST, MST, SST, i.e. large, medium and small, as defined with respect to their collecting mirror size) will have different field-of-views and have been designed to access different energy regimes from 20 GeV up to 300 TeV. Each site will be equipped with 4 LSTs and 15 MSTs (25 at the southern site) sensitive to the energy range from 20 GeV up to 10 TeV; the southern site will also be equipped, in addition, with 70 SSTs designed to extend the observation window from few TeV up to 300 TeV, and spread out over several square kilometers. The optics of the telescopes are diverse: both single and double mirror configurations are used with parabolic, Davies-Cotton or Schwarzschild-Couder design, the latter only for the double mirror telescopes. Additionally, the cameras at the focal plane of the telescopes will be equipped with either photomultiplier tubes or silicon photomultipliers as well as with custom-made electronics. Given this diversity of telescopes as well as their placement on each site, an overall coherent calibration of the full array is a challenging task that needs to take into account different sources of uncertainties together with the peculiarities of each telescope. In brief, tools are needed to inter-calibrate telescopes of a same class and cross-calibrate sub-array of telescopes of different class, without forgetting the absolute calibration of each single telescope. The final aim is to achieve correct and accurate calibration coefficients and their evolution in time to be used for the analysis of the observational data.

Several principles must be fulfilled for a successful array calibration strategy; among them:

- calibration activities should not (or only minimally) interfere with regular data taking;
- sources of calibration that come for free should be exploited as much as possible, such as muons, electrons and protons, and the air shower data taken by the telescopes;
- calibration should provide redundancy and possibility of cross-checking and estimate the scale of systematic uncertainties;
- calibration equipment must be robust, reliable and require minimal maintenance;
- calibration equipment and methods must be fully integrated in the CTA array control systems and all automatized data analysis procedures;
- the cost of calibration activities (both in terms of investment as human resources) must be in relation to the achieved improvement in precision, accuracy or recovered amount of observation time or data.

Furthermore, the requirements imposed by the CTA on calibration accuracy to achieve its science goals are very challenging. As an example, the systematic error with which the energy scale is determined must not exceed 10% which is a much stronger requirement than that achieved by any previous IACT experiment. At the same time, current IACTs are sensitive only to a part of the energy spectrum covered by the CTA, whereas calibration becomes more and more difficult towards the lower and upper ends of the energy range, either due to higher and higher atmospheric shower heights and increased intrinsic shower fluctuations, or due to less and less event statistics and large amount of light captured in individual pixels where saturation effects start to play a role. Finally, the projected lifetime of the CTA will be a factor 3 larger than that of current IACTs, and long-term degradation effects will need a more profound understanding. Successful calibration of the CTA therefore requires methods and instruments outperforming the baseline established by the current IACTs [3].

In this contribution we specifically refer to the CTA array calibration and present those methods that, applied directly to the acquired observational CTA data, will ensure that the calibration is correctly performed to the stringent precision required by the CTA, and the calibration equipment that, external to the telescopes, is currently under development and testing. Last but not least, designing an overall calibration strategy for the CTA is a cross-activity, which builds upon a close interaction between the different telescope developers and groups involved in atmospheric monitoring, array control, central scheduler, data analysis and Monte Carlo simulation. Some of the related interactions are briefly described in the operative array calibration procedure sketched in the last section of this contribution.

## 2. CTA Array Calibration – Methods and Equipment

The array calibration can be accomplished by applying a set of tools (methods and equipment) that will provide the update of characteristic parameters, necessary to ensure that the Observatory fulfills all high-level performance requirements. Some of this updates will be performed on-site and on-time, but others will be applied off-site due to, for example, the specific method, to the needed level of statistics or to fine corrections to be applied after periodic calibrations.

### 2.1 Calibration using observational CTA data

The array calibration methods using the direct observational CTA data make use of the analysis of muon rings and of cosmic-ray air shower images; additionally, the analysis of the reconstructed electron spectrum and the possible use of the Cherenkov Transparency Coefficient are considered. The main outcomes are the calibration of the telescopes' optical throughput, monitoring of their point spread function and cross-calibration of the telescope response efficiencies throughout their covered energy range.

The muon calibration method has been applied to practically all previous IACTs and is based on the analysis of muon ring images. Highly energetic muons penetrating the atmosphere produce Cherenkov light that is imaged as characteristic arcs or rings by IACTs, if their impact point is sufficiently close to the telescope optical axis. The analysis of such local muon rings is a powerful and precise method to calibrate the optical throughput of any existing IACT system. In fact, the muon aperture (for which all the telescope elements are taken into account) can be analytically derived from the geometrical parameters of the detected ring image. Comparing the image charge recorded in photoelectrons to the theoretical expectation for the same ring geometry enables the overall telescope optical throughput to the light spectrum produced by muons. A dedicated study made by the CTA Calibration team [4] established the feasibility of using muon ring images as calibrators for all CTA telescopes, improving the technique currently applied and with particular attention being paid to the SSTs, present for the first time in the current panorama of the IACT systems and particularly prone to detection biases due to their small mirror area. Due to their close distances to each other and contrary to the other telescope types, LSTs will be able to detect local muon images efficiently, even with a stereo trigger imposed, and such stereo muon events will be passed directly to the central CTA array data acquisition pipeline to be analyzed. MSTs, in turn, suffer from the large shadow to the muon light, that is produced by their quadratic 8° field-of-view cameras. For the MST and particularly for the SST telescopes the stereo muon rate will tend to zero, given their smaller mirror area and larger inter-telescope distance, thus largely increasing the time within which a statistically useful number of images can be obtained [5, 6, 7]. These telescopes must hence rely on mono muon triggers and readout. Some SSTs might even need a dedicated muon trigger scheme in order to achieve the requirement of efficient trigger and flagging during science data taking. In order to keep the mono telescope data rate at an acceptable level, some ring reconstruction and characterization may be required already at the level of the camera server [8]. In order to achieve the required accuracy, the differences between the local muon and distant gamma-ray spectra need to be taken into account which require some assessment of the chromaticity of degradation of the optical elements of each telescope, a task left for external devices (see later). During each night of regular data taking, a sufficient number of muon events, apt for the calibration purposes, shall be collected along with normal science data. All in all, this method is expected to achieve about 4% accuracy if all these details are taken correctly into account.

The second baseline calibration method is devoted to the cross-calibration of telescope response efficiencies through the use of cosmic ray images. The method, as developed in the H.E.S.S. experiment [9], uses standard shower events which have the same reconstructed impact parameter between two chosen telescopes and evaluates, through specific asymmetry parameters, the reconstructed size, or energy, of such shower images, registered simultaneously in both telescopes. In case of inter-calibration (telescopes of the same class), the asymmetry parameter is defined in relation to the recorded image size and the reconstructed core distance [9]. In case of cross-calibration (telescopes of different class), the image size varies in a nonlinear way with mirror area, differing telescope hardware specifications and variation in camera design. To extend the method to cross-calibration, the asymmetry parameter analyzed is the reconstructed energy per telescope as a proxy for image size [10]. Averaged over many events, the combination of asymmetry parameters for each triggered telescope and reconstructed shower energy (the same for the triggered telescopes) is interpreted as being due to variation in telescope optical efficiencies. Final outcomes are the relative optical efficiencies of all telescopes pairs in the array, depending on their image size or reconstructed energy. Complementary to the muon calibration, the cross-calibration procedure enables long term monitoring of telescope-wise optical throughput scans through different shower size regimes; if applied to data taken under various zenith angles and observing conditions, the method would help to reduce possible angular or size-dependent effects of the response coefficients.

Two further methods are under investigation for the CTA array calibration purposes, namely the reverted use of the Cherenkov Transparency Coefficient (CTC) and the analysis of the reconstructed cosmic-ray electron spectrum.

The CTC method, developed within the H.E.S.S. experiment, quantifies the average atmospheric transparency experienced by Cherenkov light from hadronic showers ascertained from the trigger rates simultaneously read out by the telescopes, and the optical efficiency measured with the muons. This quantity is designed to be as hardware-independent as possible in order to separate hardware-related effects from those caused by large-scale atmospheric extinction [11]. It follows that, provided that atmospheric conditions over the telescopes array are uniform, the CTC may be reverted, in the sense that observational data of parts of the array are used to measure the transparency coefficient while that coefficient is then used to obtain a relative calibration of the optical throughput for another part of the array. Details and first results of a feasibility study for extension of the reverted CTC concept in CTA are described in a separate contribution to this conference [12].

High level data calibration can be performed using the cosmic-ray electron spectrum; it is a particularly promising target as the spectrum contains a strong feature (an energy break at around 1 TeV) useful to isolate systematic scaling in the energy reconstruction of the array from systematic scaling of the effective area [13]. The electron spectrum can be extracted from almost all potential CTA extragalactic observations and then compared with a strongly data selected and high statistics spectrum to check for compatibility. If an incompatibility is found in either the energy scaling or effective area it may indicate issues in the array calibration or hardware, or the global understanding of the atmosphere above the CTA, and scaling of the high level instrument response functions could be applied to correct for this. The technique can be used as complementary to the several absolute calibration techniques foreseen in CTA and as a high-level check of the health of the array (and its telescopes) over the expected 30 year operational lifetime.

Eventually, given the tremendously increased effective area of the CTA, very often known gamma-ray sources will be found in the observed science target field-of-view, which can be used for direct monitoring of sensitivity, gamma-ray point-spread function and pointing accuracy (in the case of point sources). Planned observations for sources visible by both the Northern and Southern array of the CTA in turn, will allow to cross-calibrate the instrument response functions, which strongly depend on the zenith angle of observation.

### 2.2 Calibration using non-CTA data

The CTA array calibration will benefit from methods using external measurements for the cross-calibration of the absolute energy scale with space detectors and with archival data. These methods are under evaluation and will be described in a near future.

### 2.3 Calibration equipment

The array calibration methods described above will be complemented with the use of calibrated light sources located on ground or on board unmanned aerial vehicles.

The first of such equipment is the so-called Illuminator, a device designed for measuring the detailed response efficiencies of each telescope with respect to incident wavelength, light intensity and pulse width. It will accurately monitor such behavior and its long-term degradation over the years. A detailed description of the Illuminator and of its capabilities is given in [14]. In brief, the Illuminator is a portable ground-based device, remotely controlled, designed to uniformly illuminate the telescope aperture with a pulsed or continuous reference photon flux whose absolute intensity is monitored by a NIST-calibrated photodiode. The Illuminator is placed at a certain distance from the telescope and at an elevation such that the telescope axis is directly pointing towards it. Using different illumination features (wavelength, intensity, pulse length) as well as changing the telescope pointing towards the Illuminator, several calibration purposes can be accomplished, even with respect to off-axis angles. Among them, the overall telescope spectral response (including mirror reflectivity and detector efficiency) can be measured to obtain, for each camera pixel (or for each group of few pixels), the corrective factor for the flat fielding of the camera. One after the other, all telescopes in the whole CTA array will be calibrated against the same reference light source and configuration. Moreover, as in the ASTRI SST-2M prototype absolute calibration strategy [14] for which purpose the first Illuminator has been developed, this instrument can be accompanied by a small device used as a further reference. Details of such a device, named UVscope, are given in [15]. In brief, UVscope is a well-tested stand-alone portable NIST-calibrated multi-pixels photon detector, remotely controlled, designed to measure, in single photon counting mode, light flux with spectral acceptance matching the ones of the CTA cameras. Several kinds of measurements can be performed; as an example, the calibration among telescopes can be obtained by comparing simultaneous measurements of a bright star using it as reference: all telescopes are illuminated by the same star flux, independently from their position and atmospheric attenuation. In the same way, when placed near and co-axial to the telescope receiving the light from the Illuminator, the UVscope will monitor the flux reaching the telescope aperture. This allows to completely eliminate the uncertainty due to the atmospheric transmission and propagation, hence reaching an absolute telescope end-to-end calibration. The Illuminator should be used any time a telescope's complete calibration is required, either at the telescope(s) commissioning time, or if significant change or degradation of the telescope(s) response is suspected.

Another piece of equipment, currently under a feasibility study, foresees the use of calibrated light sources located on board of unmanned aerial vehicles (UAV) that will allow the state of all telescopes to be assessed on short notice and within very short time scales, of the order of minutes. It is foreseen to be useful after events which raise serious questions about the state of the array and require a fast assessment of its normal sensitivity, such as after strong dust intrusions, earthquakes, or periods of heavy rain. Details of such a device are given in [16]. Its prototype, currently known as Octocopter, due to the geometry of its flight platform, is radio and GNSS-controlled and, equipped with light sources of selectable wavelength ranges, it can fly to heights from hundred meters to one kilometer above the array. The Octocopter can be operated during clear nights with stratified low atmospheric dust content and low wind. Under such conditions, the Octocopter stabilizes its position at a certain height while its flasher emits isotropic light pulses of nanosecond duration. Pointing the telescopes towards the Octocopter, all telescope camera parts may be scanned quickly, such allowing the cross-calibration of the optical throughput of the entire array of telescopes at different light intensities and wavelength ranges.

### 3. Procedures and Use Cases

The array calibration strategy is a cross-activity whose procedures require a close interaction between several different components: telescopes, auxiliary calibration systems, atmospheric monitoring, array control, central scheduler, data analysis and archiving and Monte Carlo simulation. In the CTA framework the procedures are defined in terms of Use Cases, a list of steps which describe a scenario as a workflow of related interactions between a user (or more generally, an "actor") and a system to achieve a goal [17]. The various steps correspond in some sense to the human questions schematized in Fig.1. Starting from an overall top level, more and more detailed sub-level Use Cases are then defined in a hierarchical tree structure that, at the same time, must take into account all boundary conditions necessary to perform the given action. Just as an example, to perform on-site test array calibration (top level use case) the "main success scenario" could include a dedicated muon run and/or the observation of a reference source. The acquired data will be then archived and a fast analysis performed on them. Obviously, there will be "pre-conditions" to be fulfilled, as for example the environmental conditions which must be within the required limits. Based on the results obtained with the fast analysis on-line, the "actor" (the Observatory, in this case) can take a decision about whether to continue with the planned observations (as "post-condition" the array results calibrated at sufficient level) or not, or modify the observation schedule. The request for this overall use case could be "triggered" by different motivations; among them, a period of very strong winds or strong dust intrusions, or whenever any inconsistency has been found in the data during the on-site analysis of the previous night, or if considered necessary after manual interventions on a telescope. Any item described before will be defined in more and more specific sub-level use cases influencing the various CTA components.

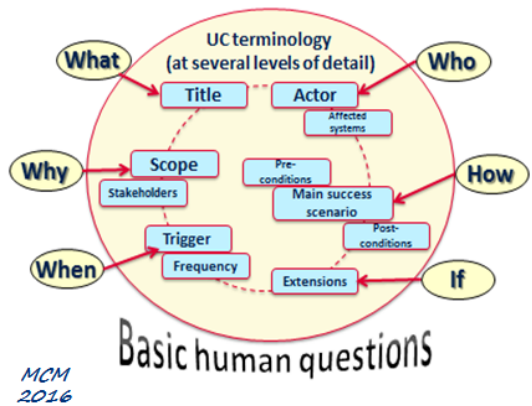

Fig.1 – Basic comparison between human and Use Case terminology (left) and (right) very simplified representation of the 'perfom on-site array calibration' top level use case (see text).

### 4. Conclusions

The calibration in the CTA framework involves several aspects: from the relative calibration of each camera [18], to the corrections for atmospheric conditions [19, 20], to the corrections for the effective pointing of all telescopes [21], to the overall array calibration. The last item has been presented in this contribution such as currently foreseen for each CTA site. Some tools are state-of-the-art in the current IACT systems, others have been specifically designed or optimized taking into account the CTA peculiarities and expectations. Altogether, such tools fulfill the requirements for a successful CTA array calibration strategy.

### Acknowledgments

This work was conducted on the context of the CTA Consortium. We gratefully acknowledge support from the agencies and organizations listed under Funding Agencies at this website: http://www.cta-observatory.org/.